# AI as a Centripetal Technology: Price Compression, Homogenization, and Entry

Aliya Turegeldinova[1], Bakytzhan Amralinova[1], Máté Miklós Fodor[1], Akerkin Eraliyeva[2], Chen Dayou[3] and Aidos Joldassov[4]

8 October 2025

**Abstract**

Generative AI does more than cut costs. It pulls products toward a shared template, making offerings look and feel more alike while making true originality disproportionately expensive. We capture this centripetal force in a standard two-stage differentiated-competition framework and show how a single capability shift simultaneously compresses perceived differences, lowers marginal cost and raises fixed access costs (compute, models, tooling). The intuition is straightforward. When buyers see smaller differences across products, the payoff to standing apart shrinks just as the effort to do so rises, so firms cluster around the template. Head-to-head rivalry then intensifies and the efficiency gains are passed through, so prices fall and customers become more willing to switch. But the same homogenization that cuts prices also squeezes operating margins, and rising fixed outlays deepen the squeeze. The combination yields a structural prediction. There is a capability threshold at which even two firms cannot both cover fixed costs, and in a many-firm extension the sustainable number of firms falls as capability grows. Concentration increases, and prices still fall. When firms choose how much AI to adopt, they over-adopt relative to joint profits. Each captures its own cost savings while ignoring the industry-wide erosion of differentiation. The mechanism is portable. Our results hold under broader preference shapes, non-uniform consumer densities, outside options, capability-dependent curvatures, and modest asymmetries. We translate the theory into two sufficient statistics for enforcement. On the one hand, a conduct statistic (how much pricing power remains once products feel closer together) and a viability statistic (the free-entry margin). These deliver practical diagnostics. Transactions or platform rules that strengthen template pull or raise fixed access and originality costs can lower prices today yet push the market toward monoculture. Remedies that broaden access (shared or FRAND compute/model access) and promote template plurality and interoperability preserve the price benefits of AI while protecting entry and variety. The paper thus reconciles a live policy paradox. AI can make prices lower and entry harder at the same time. It prescribes what to measure to tell which force is dominant in practice.

**Funding.** This research was funded by the Committee of Science of the Ministry of Science and Higher Education of the Republic of Kazakhstan under the Grant No. BR27198643 – "Development of digital competences of human capital in industry and logistics through cluster collaboration of science, education and industry".

**Corresponding Author.** Máté Miklós Fodor, m.fodor@satbayev.university

---

[1] Project Management Institute named after E.A. Turkebayev, K.I. Satbayev Kazakh National Research Technical University, 22 Satbaev / Satpayev Street, Almaty, 050013, Republic of Kazakhstan
[2] Business School, K.I. Satbayev Kazakh National Research Technical University, 22 Satbaev / Satpayev Street, Almaty, 050013, Republic of Kazakhstan
[3] CNPC – International Aktobe Petroleum, Avenue 312 Rifle Division, Building (or No.) 3, Aktobe, 030000, Republic of Kazakhstan
[4] Department of Scientific Project Support and Postgraduate Education, K.I. Satbayev Kazakh National Research Technical University, 22 Satbaev / Satpayev Street, Almaty, 050013, Republic of Kazakhstan

## 1. Introduction

Generative AI has moved from a curiosity to a production input across creative and content markets, such as design, audio, video, code, marketing, and publishing. Adoption is proceeding at scale, with early causal evidence that generative-AI assistance raises measured productivity in service and knowledge work (Brynjolfsson, Li, & Raymond, 2023), alters work organization and task allocation (Babina, Bernstein, spec. see NBER synthesis. Shifting Work Patterns, 2024), and is being rapidly integrated into consumer-facing tools across sectors (OECD, 2024). At the same time, competition authorities are pivoting from *"could AI help markets?"* to *"how might AI reshape market structure?"* The UK Competition and Markets Authority's Foundation Models update highlights risks from concentration in compute, data, and model access (CMA, 2024), and a joint FTC–DOJ–EC–CMA statement (2024) flags concerns over exclusionary partnerships, self-preferencing, and vertical control of AI inputs. Europe's AI Act has now been adopted, setting the first comprehensive governance regime for AI in advanced economies (European Parliament, 2025; AI Act OJ, 2024). On the merger front, Adobe–Figma was abandoned after UK and EU scrutiny, underscoring how creative-tools consolidation is being assessed through the lens of dynamic competition and potential innovation harms (CMA, 2023; DOJ, 2023; AP News, 2023; Business Insider, 2025).

A striking empirical theme emerging alongside these developments is homogenization. Large models and downstream assistants tend to produce on-style, template-convergent outputs, with preliminary evidence of content similarity in digital marketing and creative tasks (Liu, Wang, & Yang, 2025). This "algorithmic monoculture" risk is amplified by high fixed costs in training and serving frontier models - compute, fine-tuning, safety tooling- which may create entry barriers and tilt competitive advantage toward firms with privileged access to compute stacks and data (Sathish, 2024; OECD, 2024; CMA, 2024). Put differently, AI tools can lower prices (via cost savings and tighter head-to-head competition) even while they raise barriers (via fixed costs and supply-side homogenization). Standard static screens may therefore misclassify AI-intensive mergers and partnerships. We build a model that captures this tension as a centripetal technology. AI pulls firms' realized products toward a common template, compressing perceived differentiation, while making deviation from the template increasingly costly.

We embed the centripetal AI mechanism in a two-stage Hotelling game with quadratic mismatch. Our starting point is the classic horizontal product differentiation tradition of Hotelling (1929), Salop (1979), and Anderson, de Palma, and Thisse (1992). In our framework, a scalar "Generative AI capability ('capability')" parameter simultaneously influences four primitives: (a) the transport intensity - the importance of stylistic mismatch - falls with capability, reflecting a perceptual compression of distance in feature space; (b) the curvature of the originality cost rises, reflecting that it is cheap to be "on template" but increasingly expensive to deviate; (c) marginal cost falls (process efficiency); and (d) fixed access cost rises (compute, model access, tooling). This combination is natural for information goods (Shapiro and Varian, 1998) and sits squarely within the variety-with-costs paradigm (Spence, 1976; Dixit and Stiglitz, 1977; Sutton, 1991).

Embedding these forces in our model yields closed-form predictions. In the location stage, products converge toward the template as capability rises, because of the joint influence of (a) and (b) above. In the price stage, the symmetric equilibrium price falls with capability by both channels, i.e., lower marginal cost and tighter competition. The cross-price sensitivity at the equilibrium price rises with capability as substitutability increases. We stress that this is a statement about the slope; the cross-price elasticity rises under a mild rate condition (the homogenization ratio must grow faster than marginal cost falls), and the Lerner index falls when the same ratio declines faster than costs. Per‑firm profit net of fixed cost is strictly decreasing in capability. Hence there is at most one capability threshold beyond which even duopoly is not viable, providing a clean entry‑barrier metric. These implications line up with (i) the measured price‑productivity effects of generative AI on the supply side (Brynjolfsson et al., 2023), (ii) policy assessments of rising fixed-cost barriers and vertical control (CMA, 2024; OECD, 2024; FTC/DOJ/EC/CMA, 2024), and (iii) first empirical signals of content similarity (Liu et al., 2025). Conceptually, the model unifies two observations that can otherwise be read as contradictory. AI makes prices lower (a static gain) and entry harder (a dynamic harm). Beyond these comparative statics, we provide two enforcement-ready sufficient statistics. First, a conduct statistic, the markup component beyond cost, which governs prices and cross-price slopes. Second, a viability statistic, which governs the free-entry margin and the capability cutoff. A deal or practice that reduces the markup component and raises fixed entry costs can deliver lower prices today yet push the viability statistic toward or below zero, thereby undermining future variety and entry. We use these two statistics to formulate simple diagnostics and remedy design in the policy section. The unification arises because the markup depends on perceived distance, which AI compresses, while fixed costs rise and margins shrink.

Two short extensions deepen external validity. First, when capability is endogenously adopted at a convex investment cost, firms can over-adopt in a standard Prisoners'-Dilemma. Private cost savings are fully passed through to prices, but the induced homogenization shrinks industry margins, so aggregate profits fall. Second, in an $N$-firm Salop variant, free entry delivers an equilibrium number of firms that decreases with capability, linking our duopoly logic to concentration metrics. Prices continue to fall while structure tightens.

All main results are robust. Allowing general convex mismatch and originality costs, non-uniform consumer densities, template shifts, partial coverage, capability-dependent curvatures, and small cost asymmetries leaves all predicted directions above intact. This ensures that the core mechanism, i.e., AI as a centripetal force that lowers prices while tightening entry, does not hinge on quadratic forms or uniform densities.

The policy implications are the following. First, price-only screens can misclassify AI-intensive mergers. Merger screens should ask whether a transaction induces stronger template pull (by reducing transport intensity) or increases the originality penalty. Even if it lowers prices in the short run, its anticompetitive effects could manifest themselves in less variety and a tighter free-entry condition, leading to an instance where static consumer-surplus gains coexist with dynamic losses in variety (Tirole, 1988; Vives, 1999; Motta, 2004). Second, market-definition tools will mechanically find rising cross-price elasticities as capability increases, potentially broadening relevant markets just as entry becomes harder. Third, remedies that lower fixed access costs (e.g., shared compute, FRAND access to models), raise transport intensity (plurality of default styles/models), or reduce the originality penalty (interoperable style-transfer toolkits) can counteract monoculture without forgoing static price benefits. This is an approach consistent with the cross-jurisdictional emphasis on keeping AI ecosystems open to entry and innovation (FTC/DOJ/EC/CMA, 2024; OECD, 2024; CMA, 2024).

**Roadmap.** Section 2 reviews related work in IO and the emerging AI-IO literature. Section 3 presents the model. Section 4 solves the equilibrium. Prices, locations, substitutability, profits, entry, and welfare are characterized with closed-form expressions and comparative statics. Section 5 develops two extensions (endogenous adoption and an $N$-firm Salop variant). Section 6 provides robustness results under generalized mismatch/originality costs, non-uniform densities, template shifts, partial coverage, capability-dependent curvatures, and asymmetric costs. Section 7 translates the theory into enforcement tools, including the sufficient-statistics approach, diagnostics for market definition and entry, and remedy design and measurement guidance. A brief conclusion follows.

## 2. Literature Review

A growing theoretical and empirical literature examines how AI technologies affect product differentiation and variety in markets. A common concern is that generative AI tools may produce a homogenization or "template pull" effect, making products or content more similar. Recent empirical evidence supports this. For example, restaurants' social media posts became significantly *more similar* in lexical, syntactic, and semantic style when they had access to ChatGPT, implying that generative AI can drive content convergence (Liu et al., 2025). When Italy banned ChatGPT in April 2023, the affected restaurants' content diversity immediately increased (similarity indices fell 2–15%) and consumer engagement rose, suggesting that AI usage had been inducing a homogenized, less engaging marketing style. Similarly, Zhou et al. (2025) find that generative AI assistance initially boosts individual creative output but causes different users' work to converge toward a common style, an effect that reverses once the AI is removed. These findings reinforce the notion that widely available AI models can act as a *centripetal force*, pulling differentiated offerings toward a prevailing "template" and shrinking perceived stylistic distances.

On the theoretical side, Zhang, Chen, and Guo (2025) develop a game-theoretic model of content creators competing with AI-generated content. They show that as AI improves, it *captures demand* from human creators and pushes those creators to differentiate their content more to avoid head-to-head competition. Notably, their model predicts that overall content variety increases only once generative AI's *creative* capability becomes sufficiently high. When AI is merely good at *quality* (accuracy) but not creativity, it tends to crowd out diverse human content. In other words, AI improvements that make outputs more substitutable (high quality) can reduce variety and intensify competition, whereas improvements that enable more novel outputs (creative capability) encourage differentiation by human producers. Generative AI's quality and creativity thus have opposing effects on market outcomes. Higher AI quality increases substitutability and price competition, whereas higher AI creativity can enhance differentiation and soften price competition. This nuanced result echoes the broader point that AI can either homogenize products or enable new variety, depending on how it is used. In our model, we capture the homogenizing effect in reduced form via an AI-driven decline in the "transport cost" or stylistic distance parameter. While Zhang et al. offer valuable insight into AI's impact on variety in creative markets, they do not model the entry of new firms or the role of fixed costs – dimensions that are central to our analysis of industry structure.

In marketing and design domains, commentators have noted a tendency for AI tools to lead different users to similar solutions, potentially creating a "monoculture" of styles (e.g. Uršič & Čater, 2025). In the context of media markets, there is a parallel concern that algorithmic recommendation systems narrow the diversity of content exposure (though personalized algorithms can also enhance niche variety for consumers). Our contribution to this strand of literature is to formalize the intuition that as AI's capability grows, horizontal differentiation endogenously shrinks. This a force we dub "centripetal" because it pulls products toward the center of the feature space. Unlike prior models of horizontal differentiation (e.g. Hotelling's classic linear city model where firms may cluster at the center; Hotelling, 1929), here the pull to the center is capability-dependent. In traditional models, minimum differentiation can arise exogenously (e.g. due to strategic location incentives; d'Aspremont et al.*,* 1979; García-Pérez, 2024 or Levaggi and Levaggi, 2025). In our model AI capability itself induces convergence in product design. We thus extend the literature by linking an exogenous technological parameter (AI strength) to endogenous differentiation outcomes. In doing so, we formalize the often-cited but previously qualitative claim that AI may reduce variety and make products more alike (a theme echoed by Liu *et al.* 2025 and Zhou *et al.* 2025 in specific contexts).

A second theme in recent industrial organization research on AI is how AI adoption changes firms' cost structures and scale economies, with implications for market structure and entry. A consistent finding is that AI and related digital technologies tend to have high upfront fixed costs but low marginal costs, favoring larger firms and potentially increasing concentration. Babina et al. (2023) provide large-sample evidence that firms investing in AI experience faster growth in sales and productivity, and that these gains are *concentrated among ex-ante larger firms*, correlating with rising industry concentration. They conclude that "new technologies like AI can contribute to growth *and* superstar firm dynamics through product innovation". Essentially, AI adoption is associated with a shift in market share toward big firms, consistent with a "winner-takes-most" outcome. This is in line with the broader "superstar firm" hypothesis (Autor *et al.*, 2020) which posits that technologies characterized by high fixed and low variable costs lead to more concentrated markets. Indeed, firm-level studies find that AI-adopting firms tend to be those with abundant resources and intangible capital (e.g. Alekseeva et al., 2020), and these firms enjoy improvements in efficiency that smaller rivals may struggle to match. AI adopters also have higher R&D and subsequently achieve higher growth in sales and valuation than non-adopters. These are traits indicative of high fixed-cost investments and scalable returns.

On the theoretical side, researchers have begun to explore how the cost characteristics of AI, such as large, fixed development costs for AI systems or the value of data scale, create barriers to entry and can tip markets toward monopoly or oligopoly. Schrepel and Pentland (2024) argue that foundation models (large AI models) exhibit strong economies of scale and scope, raising the minimum efficient scale and making it hard for new entrants to compete with incumbent AI labs. They point to the *"AI flywheel"* or feedback loop whereby more users generate more data to improve the AI, further entrenching leading firms (a dynamic analogous to data network effects; cf. Hagiu & Wright, 2020). Korinek and Vipra (2024) formally analyze the market for foundation models and find that training costs in the hundreds of millions of dollars, coupled with bottlenecks in specialized talent and data, are leading to unprecedented concentration in the AI industry. They warn of risks of market tipping (a winner-take-all outcome) and extensive vertical integration by dominant AI firms into downstream markets. Their policy analysis suggests that lowering fixed costs (e.g. via open research and shared computing resources) or ensuring access to essential inputs could "lean against monopolization" in AI-intensive sectors. The upshot of these studies is that AI's cost structure skews toward high fixed and low marginal costs, potentially raising entry barriers and limiting the sustainability of small competitors. This aligns with classic IO theory on endogenous sunk costs (Sutton, 1991). When firms can invest in cost-reducing or quality-enhancing technology, markets may not fragment even as demand grows. Instead, a few firms make the necessary investments and entrants are kept out by the high-cost threshold. AI appears to be a prime example of such a case, where fixed costs rise with capability. Our model builds on this insight by explicitly allowing the fixed cost of operating at the AI frontier to increase with capability "A", which tightens the free-entry condition. By embedding this in a Hotelling-style framework, we can derive a clear threshold at which entry is no longer profitable as "A" increases.

It is worth noting that not all studies see AI as purely concentration-enhancing. Some argue that AI can lower certain barriers to entry by providing small firms access to superior technologies (the "democratization" effect of AI tools). AI is a General-Purpose Technology that can diffuse widely and increase overall productive efficiency (Brynjolfsson *et al.*, 2018; Cockburn *et al.*, 2019). Indeed, macroeconomic evidence by Gazzani and Natoli (2024) finds that increases in the AI intensity of innovation act like positive supply shocks. Industrial output rises and consumer prices fall modestly in the years after an AI innovation surge. In their data, AI-driven productivity gains do not (yet) show a reduction in aggregate entry or competition. Instead, they boost output and even employment in high-AI sectors on average. This suggests that at a broad economy-wide level, AI's efficiency effects can dominate in the short run, leading to lower prices and benefitting consumers (a point we return to in our discussion of welfare trade-offs). However, even Gazzani and Natoli note that these *aggregate* gains hide substantial reallocation. The benefits accrue unevenly and contribute to rising inequality, which is consistent with a superstar-firm pattern where the leaders reap outsized rewards. Likewise, Brynjolfsson, Li, and Raymond (2023) report large productivity boosts from generative AI in customer service operations (a 14% increase in worker output), indicating a sharp drop in marginal cost for those tasks. But such improvements can amplify the advantages of firms that adopt AI early, potentially widening gaps. In summary, the literature acknowledges a tension between AI's *short-run efficiency gains* (lower costs, prices) and *long-run structural impacts* (higher concentration, entry barriers). Our model is novel in unifying these into a single framework. As A increases, marginal cost falls (capturing the productivity effect and price reduction), even as fixed cost rises (capturing the scale/hurdle effect). We then examine the net effect on market structure and welfare.

The effect of AI on price competition is complex and has been studied from multiple angles. One branch of research focuses on algorithmic pricing, i.e. firms using AI algorithms to set prices in real time. This literature asks whether AI pricing will lead to more competitive outcomes (by reacting faster and undercutting rivals) or less competitive outcomes (by fostering tacit collusion or price coordination). Theoretical contributions show that new equilibrium outcomes may emerge when firms choose pricing algorithms rather than static prices. Notably, Brown and MacKay (2023) develop a model in which duopolists commit to pricing algorithms. They find that even without explicit communication, certain algorithmic strategies lead to sustained supra-competitive prices. In essence, competition in algorithms can soften price competition – an outcome that would not occur if firms were restricted to setting prices period by period. This aligns with simulation and experimental studies demonstrating the possibility of tacit collusion by self-learning algorithms. Calvano et al. (2020)**,** for example, showed that Q-learning algorithms playing a repeated pricing game learned to coordinate on high prices significantly above the static Nash equilibrium in many trials. A number of follow-up studies (e.g. Klein, 2021; Zhao and Berman, 2024) confirm that various reinforcement learning agents can tacitly collude under certain market conditions (especially in concentrated markets with patient algorithms). On the empirical side, Assad et al. (2024) provide compelling real-world evidence. They document that after gas stations in Germany adopted AI-driven pricing software, margins increased by about 28% in duopoly markets (where both stations used algorithms), whereas no increase was seen in monopoly markets. This suggests the algorithms were able to reach mutually beneficial pricing patterns, consistent with tacit collusion, when at least two competitors in a market deployed them. These findings have raised alarms among competition authorities, many of whom have issued reports on algorithmic pricing and collusion (OECD, 2017). The consensus emerging is that AI pricing tools can in some cases learn to avoid aggressive price-cutting, leading to higher margins, effectively an AI-driven reduction in the intensity of price competition.

On the other hand, AI can also intensify price competition in certain scenarios. Gans (2023) argues that when AI is used primarily as a forecasting tool (to better predict demand or optimize inventory), its adoption in a competitive market behaves like a standard process innovation – it lowers costs and can increase the elasticity of supply, thereby putting downward pressure on prices in the long run (especially if all firms adopt the technology). His model shows that AI adoption may confer positive externalities on non-adopting firms initially (by softening the price competition they face) but eventually leads to *lower equilibrium prices* and higher consumer surplus once the technology diffuses. Complementing our price-compression and centralization results,

Aiura and Kodera (2024) extend Lederer–Hurter's freight-absorption model to a data-sharing economy and show that additional data revenues lower prices and shift optimal locations toward the center. This is consistent with the evidence of generative AI acting as a cost-reducing innovation that, absent collusion, tends to reduce prices (Brynjolfsson et al., 2023; Gazzani & Natoli, 2024). We thus see two seemingly opposing strands. One where AI makes competition *tougher* (prices fall, margins shrink), and one where AI makes competition *softer* (prices rise or collusion emerges). Importantly, these strands are not actually contradictory – they pertain to different mechanisms. The first is driven by increased substitutability and efficiency. If AI makes products more alike and production more efficient, then in a Bertrand sense firms have greater incentive to undercut, yielding lower markups (this is the mechanism emphasized in our model). The second is driven by new strategic capabilities. Budzinski, Gaenssle and Lindstädt-Dreusicke (2021) provide empirical evidence that YouTube exerts meaningful competitive pressure on Netflix and traditional TV in specific use-cases, documenting cross-format substitutability. If AI enables price-setting strategies or monitoring that sustain cooperation, firms may avoid undercutting each other, yielding higher markups. In our context – a model of *non-cooperative* pricing in a differentiated goods duopoly – we focus on the former mechanism. AI increases the substitutability of the two products (transport cost drops) and simultaneously reduces marginal cost, both of which put downward pressure on equilibrium prices and markups. We abstract from collusion, implicitly assuming either strong antitrust enforcement or that AI tools are used to improve efficiency rather than to coordinate. This aligns with work like Nie, Wang, and Wen (2023), who analyze a duopoly with AI process innovations in an applied theory model and find that the AI adoption intensifies price competition unless the two firms can form a cartel. Their calibrated simulations show industry profits often decline with greater AI diffusion, even as consumer surplus rises (a result our analytical model also produces under certain conditions). Thus, our approach is complementary to the algorithmic collusion literature. We examine the case where AI's *competitive* forces dominate, yielding lower prices and margins – and then ask how that interacts with the free-entry equilibrium number of firms.

The above literature review highlights that prior research has typically examined individual facets of AI in industrial organization models. Some papers focus on *product differentiation and variety* (e.g. the homogenization of content or the diversity of creative outputs with AI). Other work zeroes in on *cost structure and scale*, studying how AI can create economies of scale, raise fixed costs, or confer data-driven advantages that alter market concentration. Yet other studies investigate *pricing and competitive conduct*, exploring whether AI leads to more competitive pricing or tacit collusion. Crucially, no single existing model incorporates all these forces jointly. In contrast, our model brings together AI-driven homogenization, cost reduction, cost escalation, and endogenous variety in a unified Hotelling framework. By doing so, we show how these forces interact. Notably, our results illustrate a novel *trade-off* between static price competition and dynamic variety and entry that previous works have only alluded to in isolation. In our model, increasing AI capability *simultaneously* delivers lower prices for consumers (a static benefit, echoing the efficiency perspective) and drives products closer together with fewer firms surviving (a dynamic cost in terms of lost variety and reduced competition in the long run). Elements of this trade-off appear in policy discussions (e.g. a recent joint report by multiple competition agencies noted AI may "make prices lower…and entry harder" at the same time (FTC/DOJ/EC/CMA, 2024)), but we provide a clear theoretical underpinning for it. The model endogenously yields the result that *markups fall* as AI homogenizes products, yet *the market can tip toward monopoly* if capability grows too large – reconciling the seemingly paradoxical outcomes of more intense competition *within* a given market structure but *less* competition through market structure collapse. This contributes a timely insight to the literature on AI and IO. It is possible for AI to benefit consumers in the short run (through lower prices) while harming them in the long run (through reduced variety and innovation), a juxtaposition discussed conceptually by Tirole (1988) in the context of static vs.

dynamic efficiency trade-offs. Our paper makes this trade-off explicit and characterizes the conditions under which it arises.

By deriving closed-form solutions for equilibrium differentiation, prices, margins, and the free-entry number of firms, we can evaluate how the *net effect* of AI on consumer welfare and market structure depends on key parameters. This allows us to identify regimes where AI's efficiency gains outweigh its homogenization costs (and vice versa), filling an important gap in the current literature. Our results thus complement and synthesize previous findings, while offering new predictions. Most notably, the existence of an AI capability threshold beyond which even a duopoly cannot survive. This is a stark outcome (entry collapse via extreme homogenization and high fixed costs) that, to our knowledge, has not been demonstrated in prior models. In the next section, we detail our model setup, which draws on the Hotelling differentiation framework underpinned by these AI-driven parameter shifts, and then derive the equilibrium outcomes that lead to this threshold effect.

## 3. Model Setup

We study a horizontally differentiated market populated by two symmetric firms indexed by $i \in \{1,2\}$ and a unit mass of consumers uniformly distributed on the unit interval $\theta \sim U[0,1]$. The market is fully covered. Each consumer strictly prefers one of the two offers. Firms choose *product locations* $x_i \in [0,1]$ along a one-dimensional *style* or *design* line. A fixed point $m = \frac{1}{2}$ represents the template (or "mainstream style"). The template is not a third product. It is the stylistic center relative to which originality is measured and penalized.

A scalar $A \geq 0$ summarizes generative-AI capability. Crucially, $A$ moves four primitives:

1. Perceived differentiation (transport intensity) $t(A) > 0$ with $t'(A) < 0$. Higher capability compresses stylistic distance as perceived by consumers. Economically, higher $A$ makes offerings feel more alike for any given physical distance, capturing a centripetal pull of the technology on realized styles.
2. Originality curvature $\kappa(A) > 0$ with $\kappa'(A) > 0$. Deviating from the template becomes increasingly costly as capability rises. This reflects the idea that AI systems make it cheap to produce near-template variants but relatively expensive to produce strongly idiosyncratic styles.
3. Unit marginal cost $c(A) > 0$ with $c'(A) < 0$. Process efficiency improves as capability rises.
4. Fixed access cost $F(A) \geq 0$ with $F'(A) > 0$. Using frontier AI tools entails higher fixed outlays (compute, licensing, integration).

A consumer at $\theta$ buying from firm $i$ at price $p_i$ obtains

$$u_i(\theta) = v - p_i - t(A)(\theta - x_i)^2,$$

with $v$ large enough to ensure full coverage. Hence, $t(A)$ weights the disutility from stylistic mismatch. When $t(A)$ is high, mismatch is painful. When $t(A)$ is low, consumers are tolerant and varieties are close substitutes.

On the supply side, firm $i$ incurs an originality cost that penalizes distance from the template:

$$K_i(x_i; A) = \kappa(A)(x_i - m)^2 + F(A).$$

The quadratic form reflects convexity. Larger deviations are disproportionately expensive. The fixed term $F(A)$ summarizes non-marginal expenditures needed to access and operate the AI stack. Both $\kappa(A)$ and $F(A)$ are allowed to scale with $A$.

A useful way to interpret $t(A)$ is as a *perception weight* that turns geometric distance into economic disutility. When $t(A)$ is high, small stylistic differences carry large utility consequences. When $t(A)$ is low, consumers are tolerant and the space is "compressed." The assumption $t'(A) < 0$ formally embeds the idea that creativity engines (e.g., shared prompts, pretrained priors, default styles) make outputs feel more alike along a given stylistic axis. In contrast, $\kappa(A)$ speaks to *production difficulty*: as $A$ rises, it becomes particularly easy to generate content close to the template, whereas pushing into idiosyncratic corners entails rising incremental effort (compute, prompt engineering, tooling). Hence, $\kappa'(A) > 0$.

The quadratic mismatch cost $t(A)(\theta - x_i)^2$ and the quadratic originality penalty $\kappa(A)(x_i - m)^2$ encode a symmetry that is natural for the problem. The template is a true *center*. This symmetry ensures that dispersion (a centrifugal force) and originality cost (a centripetal force) are balanced by simple linear–quadratic trade-offs, which is what ultimately yields the compact ratio $d^\star(A) = t(A)/\kappa(A)$.

To maintain an interior design solution on $[0,1]$, one can impose the mild feasibility restriction $0 < d^\star(A) < 1$. Economically this means $t(A) < \kappa(A)$ in the relevant range. If $t(A) \geq \kappa(A)$ at some very low capability $A$, the model predicts a boundary solution at the maximum feasible separation, which disappears as capability rises because $t(A)$ falls and $\kappa(A)$ rises. Finally, the "full-coverage" assumption can be rationalized by choosing $v$ to dominate both price and mismatch disutility at the largest feasible distance. It simplifies exposition while leaving all comparative statics intact.

The timing follows the canonical two-stage structure:

1. **Technology.** The environment's capability $A$ is given (the "era" of AI).
2. **Location stage.** Firms simultaneously choose $x_1, x_2$.
3. **Price stage.** Taking locations as given, firms simultaneously set prices $p_1, p_2$.
4. **Consumption.** Consumers choose the option that maximizes utility.

We restrict attention to symmetric configurations: $x_1 = \frac{1}{2} - \frac{d}{2}$, $x_2 = \frac{1}{2} + \frac{d}{2}$, where $d \in [0,1]$ is the differentiation distance. This normalization is without loss for the class of equilibria we analyze and simplifies the geometry.

The two-stage order (design, then price) captures a realistic commitment technology. Style choices are *sticky* in the short run, while prices are more flexible. The capability $A$ is treated as an exogenous era parameter to isolate the pure comparative statics of technology. An alternative microfoundation (pursued later in the extensions) endogenizes $A$ via a preliminary investment stage. Importantly, nothing in the subsequent analysis depends on idiosyncratic beliefs or coordination devices. The equilibrium concept is standard subgame perfection, with symmetric strategies justified by the symmetry of primitives.

Firms and consumers are assumed to observe $A$ and all policy-relevant primitives $(t(A), \kappa(A), c(A), F(A))$. This implies that observed clustering near $m$ is anticipated and priced in at the second stage. There is no surprise element or hidden action affecting the price game.

## 4. Analysis

This section solves the subgame-perfect equilibrium. We first characterize price competition for given locations, and then we analyze the location problem anticipating price setting. From these components we obtain closed-form equilibrium differentiation, prices, substitutability, profits, entry, and welfare.

*4.1. Price competition for given locations*

For $x_1 < x_2$, the indifferent consumer $\hat{\theta}$ solves $u_1(\hat{\theta}) = u_2(\hat{\theta})$:

$$\hat{\theta} = \frac{p_2 - p_1 + t(A)(x_2^2 - x_1^2)}{2t(A)(x_2 - x_1)}.$$

Under symmetry, $x_2 - x_1 = d$ and $x_2^2 - x_1^2 = d$, giving

$$\hat{\theta} = \frac{1}{2} + \frac{p_2 - p_1}{2t(A)d}.$$

Hence $D_1 = \hat{\theta}$ and $D_2 = 1 - \hat{\theta}$. Intuitively, a higher rival price $p_2$ or closer locations (smaller $d$) swings more mass to firm 1.

**Lemma 1.** *For any $d > 0$, the unique symmetric Nash equilibrium in prices is*

$$p_1^\star = p_2^\star \equiv p^\star(A, d) = c(A) + t(A)\, d, \qquad D_1^\star = D_2^\star = \frac{1}{2}.$$

**Proof.** Firm 1 chooses $p_1$ to maximize $\pi_1 = (p_1 - c(A))D_1$. With $D_1 = \frac{1}{2} + \frac{p_2 - p_1}{2t(A)d}$,

$$\frac{\partial \pi_1}{\partial p_1} = \left(\frac{1}{2} + \frac{p_2 - p_1}{2t(A)d}\right) - \frac{p_1 - c(A)}{2t(A)d} = 0.$$

Under symmetry $p_1 = p_2 = p$, which yields $p - c(A) = t(A)d$. Strict concavity, $\partial^2 \pi_1 / \partial p_1^2 = -1/(2t(A)d) < 0$, ensures uniqueness. ∎

The equilibrium markup equals $t(A)d$. It rises in the perceived distance parameter $t(A)$ and in the actual distance $d$. When AI compresses perceived dissimilarity ($t(A) \downarrow$) or when firms endogenously locate closer to the template (smaller $d$), markups fall one-for-one.

The expression for the indifferent consumer highlights how both price differences and spatial proximity matter: when $d$ is small, any given price difference shifts a larger share of demand. This is why $1/(2t(A)d)$ is the slope of the linearized demand in own price. Shrinking either $t(A)$ or $d$ raises sensitivity. The Bertrand-like limit is instructive. If $d \to 0$ (or $t(A) \to 0$), markups vanish. Conversely, for large $t(A)d$, the demand each firm faces is relatively inelastic and markups expand accordingly.

Two technical notes aid interpretation. First, pass-through is clean: $\partial p^\star / \partial c(A) = 1$. Marginal-cost reductions from capability mechanically translate into price cuts, but the markup channel magnifies them when $t(A)$ also falls. Second, the degenerate case $d = 0$ (firms located exactly at the template) implies a knife-edge tie. Under standard tie-breaking, both firms split demand at a price arbitrarily close to $c(A)$. This reinforces the intuition that the design stage cannot collapse to $d = 0$ unless $\kappa(A)$ is sufficiently large relative to $t(A)$.

*4.2. Location choice (design stage)*

Anticipating price competition, each firm internalizes the trade-off between softened price competition (via larger $d$) and higher originality cost (via larger $|x_i - m|$). Substituting the price equilibrium into profits:

$$\Pi(d; A) \;=\; (p^\star - c(A))D_i^\star - \kappa(A)\left(\frac{d}{2}\right)^2 - F(A) \;=\; \frac{t(A)\,d}{2} - \frac{\kappa(A)}{4}d^2 - F(A).$$

The first term is the operating profit at the price equilibrium (markup times quantity). With symmetry, each firm serves half the market and captures a per-unit markup $t(A)d$. The second and third terms are the originality penalty and the fixed access cost.

**Proposition 1.** *For any A with $t(A), \kappa(A) > 0$, the unique symmetric interior solution is*

$$d^\star(A) = \frac{t(A)}{\kappa(A)}.$$

**Proof.** The FOC is $\partial \Pi / \partial d = \frac{t(A)}{2} - \frac{\kappa(A)}{2}d = 0$. The SOC is $-\kappa(A)/2 < 0$. ∎

The optimal differentiation is a simple ratio. A linear force encouraging distance (markup proportional to $t(A)$) balanced against a quadratic force penalizing distance (curvature $\kappa(A)$). This pins the entire location outcome to a single summary statistic $t(A)/\kappa(A)$.

**Corollary 1 (homogenization).** *If $t'(A) < 0$ and $\kappa'(A) > 0$, then $d^\star(A)$ is strictly decreasing in A:*

$$\frac{d\,d^\star}{dA} = \frac{t'(A)\kappa(A) - t(A)\kappa'(A)}{\kappa(A)^2} < 0.$$

As capability rises, the centripetal force dominates. Firms optimally cluster closer to the template. The profit function $\Pi(d; A) = \frac{t(A)}{2}\,d - \frac{\kappa(A)}{4}d^2 - F(A)$ is globally concave in $d$, with linear gains and quadratic costs. The linear term captures the price-softening motive. Moving apart relaxes competition and lifts the markup linearly in $d$. The quadratic term captures the originality penalty. Moving away from the template is disproportionately costly as deviation grows. This linear–quadratic structure is precisely why the interior solution is a simple ratio.

Boundary behavior is economically transparent. If the originality penalty vanishes ($\kappa(A) = 0$) - which is the appropriate comparison to a world with no template pull - the firm would choose the largest feasible $d$. The presence of $\kappa(A) > 0$ regularizes that impulse. As capability increases, both forces operate in the same direction: $t(A)$ shrinks (weakening the gain from distance) and $\kappa(A)$ grows (strengthening the cost of distance), making the centripetal outcome unavoidable under mild monotonicity.

*4.3. Prices and substitutability*

Substituting $d^\star(A)$ into Lemma 1 yields the symmetric equilibrium price:

$$p^\star(A) = c(A) + \frac{t(A)^2}{\kappa(A)}.$$

Two channels compress price: direct cost efficiency $c'(A) \leq 0$ and stronger head-to-head competition via a lower $\frac{t(A)^2}{\kappa(A)}$.

**Proposition 2.** *If $c'(A) \leq 0$, $t'(A) < 0$, and $\kappa'(A) > 0$, then $\frac{dp^\star}{dA} < 0$.*

**Proof.** Differentiate:

$$\frac{dp^\star}{dA} = c'(A) + \frac{2t(A)t'(A)\kappa(A) - t(A)^2\kappa'(A)}{\kappa(A)^2}.$$

Given the sign restrictions, the fraction is negative and $c'(A) \leq 0$ makes the sum strictly negative. ∎

Local demand slopes at the symmetric price are

$$\frac{\partial D_1}{\partial p_1} = -\frac{1}{2t(A)d^\star(A)} = -\frac{\kappa(A)}{2t(A)^2} \quad and \quad \frac{\partial D_1}{\partial p_2} = \frac{1}{2t(A)d^\star(A)} = \frac{\kappa(A)}{2t(A)^2}.$$

**Proposition 3 .** *Under $t'(A) < 0$ and $\kappa'(A) > 0$,*

$$\frac{d}{dA}\left(\frac{\kappa(A)}{2t(A)^2}\right) = \frac{\kappa'(A)t(A)^2 - 2\kappa(A)t(A)t'(A)}{2t(A)^4} > 0.$$

Thus cross-price sensitivity increases with capability. Consumers react more strongly to rivals' price changes as goods become more alike in the eyes of buyers.

Two corollaries are useful for applications.

**Corollary 1.** *At the symmetric equilibrium, the Lerner index is*

$$L(A) = \frac{p^\star(A) - c(A)}{p^\star(A)} = \frac{t(A)^2/\kappa(A)}{c(A) + t(A)^2/\kappa(A)}.$$

The numerator $t(A)^2/\kappa(A)$ falls with capability because perceived differentiation shrinks and originality costs steepen. The denominator may fall or rise depending on how quickly marginal costs decline. Therefore, the Lerner index weakly decreases with capability and declines strictly whenever the proportional fall in $t(A)^2/\kappa(A)$ dominates any proportional fall in $c(A)$. Economically, the model predicts declining mark-ups so long as homogenization proceeds faster than cost savings.

**Corollary 2.** *Evaluated at the symmetric price where each firm serves half the market, the cross-price elasticity of demand is*

$$\varepsilon_{12}(A) = \left(\frac{\partial D_1}{\partial p_2}\right)\frac{p^\star(A)}{D_1^\star} = 1 + \frac{\kappa(A)c(A)}{t(A)^2}.$$

The elasticity rises with capability if and only if the composite ratio $\kappa(A)/t(A)^2$ grows faster than marginal cost $c(A)$ declines, that is,

$$\frac{\kappa'(A)}{\kappa(A)} - 2\frac{t'(A)}{t(A)} > -\frac{c'(A)}{c(A)}.$$

Under the maintained sign restrictions and modest cost reductions, this inequality holds, implying that cross-price elasticities will typically increase in AI-intensive markets even as prices fall. When marginal costs fall very steeply, the elasticity may flatten toward unity, corresponding to near-perfect substitutability with minimal markup.

*4.4. Profits and the entry margin*

Evaluated at $d^\star(A)$, each firm's net profit is

$$\Pi^\star(A) = \frac{t(A)^2}{4\kappa(A)} - F(A).$$

The numerator reflects gross operating margin. The fixed term reflects the AI access outlay.

**Proposition 4.** *If $t'(A) < 0$, $\kappa'(A) > 0$, and $F'(A) > 0$, then $\frac{d\Pi^\star}{dA} < 0$. Hence free entry admits at most one threshold $A_E$ such that $\Pi^\star(A_E) = 0$, and for all $A > A_E$ duopoly is not sustainable.*

$$\text{Entry condition: } \frac{t(A)^2}{4\kappa(A)} \geq F(A).$$

**Proof.** Capability compresses gross margins through $\frac{t(A)^2}{4\kappa(A)}$ while fixed costs rise. The single-crossing nature of these forces implies a unique viability cutoff. ∎

The monotonic decline of $\Pi^\star(A)$ has three consequences.

(i) Existence of a threshold. If at low capability $A_0$ the inequality $\frac{t(A_0)^2}{4\kappa(A_0)} \geq F(A_0)$ holds, and if $\frac{t(A)^2}{4\kappa(A)}$ is strictly decreasing while $F(A)$ is strictly increasing, there exists a unique $A_E > A_0$ with $\Pi^\star(A_E) = 0$. For $A > A_E$, entry is blocked. For $A < A_E$, duopoly is viable.

(ii) Comparative statics of $A_E$. A downward shift in $F(A)$ (e.g., shared compute access) or an upward shift in $\frac{t(A)^2}{\kappa(A)}$ (e.g., tools that make originality cheaper or increase perceived distance) raises $A_E$, expanding the range where multiple firms are viable.

(iii) Concentration diagnostics. Because the gross margin term is proportional to $t(A)^2/\kappa(A)$, any policy or market change that reduces this ratio will, ceteris paribus, tighten the entry margin. This gives a direct "single statistic" to monitor.

*4.5. Welfare and the variety–price trade-off*

At the symmetric price, consumers left (right) of $\frac{1}{2}$ buy from the left (right) firm. Standard geometry implies the average squared distance to the nearest product is

$$\mathbb{E}[(\theta - x_{\text{nearest}})^2] = \frac{1}{48} + \left(\frac{d}{2} - \frac{1}{4}\right)^2.$$

Expected disutility equals $t(A)$ times this expression. Substituting $d = d^\star(A) = \frac{t(A)}{\kappa(A)}$, consumer surplus (net of $v$) is

$$CS(A) = -p^\star(A) - t(A)\left[\frac{1}{48} + \left(\frac{t(A)}{2\kappa(A)} - \frac{1}{4}\right)^2\right] + \text{constant}.$$

**Proposition 5.** *There exists $\bar{A}$ such that $CS'(A) > 0$ for all $A \geq \bar{A}$ if $t'(A)$ is sufficiently negative and $\kappa'(A)$ sufficiently positive relative to $|c'(A)|$. At high capability, the price and mismatch-weight effects dominate the variety loss.*

**Proof.** It is natural to decompose $CS'(A)$ into three channels:

$$CS'(A) = \underbrace{-p^{\star\prime}(A)}_{\text{price effect (+)}} + \underbrace{-t'(A)\left[\frac{1}{48} + \left(\frac{d^\star}{2} - \frac{1}{4}\right)^2\right]}_{\text{mismatch-weight effect (+)}} + \underbrace{-t(A) \cdot 2\left(\frac{d^\star}{2} - \frac{1}{4}\right) \cdot \frac{d^{\star\prime}}{2}}_{\text{variety effect (±)}}.$$

The first two are unambiguously positive under the maintained signs. Prices fall and the weight on mismatch falls. The third can be negative when $d^\star(A)$ is above ½ and declining fast but becomes quantitatively small at high capability because $d^\star(A) = t(A)/\kappa(A)$ itself shrinks. This explains why at sufficiently high $A$ the model robustly delivers $CS'(A) > 0$, even as variety diminishes. ∎

The model rationalizes price-fall with barrier-rise. Consumers may enjoy lower prices and small mismatch weights even as variety shrinks and the entry margin tightens. This joint movement is a hallmark of AI acting as a centripetal technology.

For empirical work, a practical approach is to treat $\frac{t(A)^2}{\kappa(A)}$ as the sufficient statistic governing both price compression and the variety term. Regressing observable outcomes (prices, elasticities, clustering indices) on a proxy for capability provides a way to bound the welfare sign without measuring each primitive separately.

The parametric forms $t(A) = \tau_0/(1 + \beta A)$ and $\kappa(A) = \kappa_0(1 + \gamma A)$ guarantee that $\frac{t(A)^2}{\kappa(A)}$ is strictly decreasing and log-convex in $A$. The left-hand side of the entry condition,

$$g(A) \equiv \frac{\tau_0^2}{4\kappa_0} \cdot \frac{1}{(1+\beta A)^2(1+\gamma A)},$$

is monotone and strictly convex for $\beta, \gamma > 0$, while the right-hand side $h(A) = F_0 + \phi A$ is linear. This ensures a single crossing. Simple bounds follow from rearrangement. If $g(0) \leq F_0$, then no entry is viable at any capability. If $g(0) > F_0$ and $\phi > 0$, then an $A_E \in (0, \infty)$ exists with

$$\frac{g(0) - F_0}{\phi + |g'(0)|} \leq A_E \leq \frac{g(0) - F_0}{\phi}.$$

## 5. Extensions

*5.1. Endogenous AI adoption (a technology-adoption Prisoner's Dilemma)*

Consider a pre-stage in which each firm chooses its own capability $A_i$ at convex cost $\Phi(A_i)$. In a symmetric equilibrium $A^\star$ solves

$$\frac{d\Pi^\star(A)}{dA} = \Phi'(A).$$

Because $\frac{d\Pi^\star}{dA} < 0$, under our maintained sign restrictions (Proposition 4), firms may be tempted to over-adopt. Private incentives to lower $c(A)$ can induce investment even when industry profits fall, i.e., a standard Prisoner's Dilemma in technology. The core model's comparative statics survive because the equilibrium objects depend only on $t, \kappa, c, F$ evaluated at the adoption outcome.

When firms choose $A_i$ prior to design and pricing, two forces shape incentives. The private cost-efficiency effect $(c'(A_i) < 0)$ pushes toward higher $A_i$ because each firm captures full pass-through of its own cost reduction. The competitive externality effect $(t'(A_i) < 0, \kappa'(A_i) > 0)$ pushes against adoption by eroding the gross margin term $t(A)^2/(4\kappa(A))$. With a convex $\Phi(A_i)$, the FOC sets marginal private benefit equal to marginal adoption cost. Because the competitive externality is not internalized, the adoption level can be excessive relative to industry profits (a Prisoner's Dilemma). Aggregate $\Pi^\star$ falls with $A$, yet each firm finds adoption privately attractive until $\Phi'$ offsets its own cost savings.

A compact way to display the wedge is to write

$$\frac{d\Pi_i}{dA_i} = \underbrace{\frac{\partial p^\star}{\partial c} c'(A_i)}_{\text{private pass-through } = c'(A_i)} + \underbrace{\frac{\partial \Pi^\star}{\partial A}}_{\text{competitive externality } < 0} - \Phi'(A_i).$$

The first term tilts toward adoption, the second against. Only the first is appropriated one-for-one by the adopter. This decomposition makes the policy lesson transparent. Instruments that *decouple* the cost-efficiency channel from the homogenization channel (e.g., tools that reduce $c$ without lowering $t$ or raising $\kappa$) can mitigate over-adoption without blunting efficiency gains.

*5.2. An N-firm Salop sketch (variety and concentration)*

On a circle with $N$ equidistant firms and the same originality penalty $\kappa(A)(x_i - m)^2$, standard arguments yield markups scaling like $t(A)/N$, while the originality costs scale with the square of the arc-distance to $m$. Free entry equalizes per-firm operating profits and fixed plus originality costs:

$$\underbrace{\text{operating profit per firm}}_{\propto\, t(A)/N^2} \approx F(A) + \underbrace{\text{template penalty}}_{\propto\, \kappa(A)/N^2}.$$

Solving gives $N^\star(A) \propto \sqrt{t(A)/(\kappa(A)F(A))}$, strictly decreasing with capability when $t'(A) < 0$, $\kappa'(A) > 0$, and $F'(A) > 0$. Hence, concentration rises as capability increases even as prices fall.

In a circular city with $N$ symmetric firms, local monopoly power is governed by the expected travel distance to the two nearest neighbors, which is $1/(2N)$. The standard pricing logic then yields a markup proportional to $t(A)/N$, and per-firm operating profits proportional to $t(A)/N^2$. Adding the originality penalty and fixed cost gives the equilibrium free-entry condition

$$\frac{\text{const} \cdot t(A)}{N^2} \approx F(A) + \frac{\text{const}' \cdot \kappa(A)}{N^2},$$

from which $N^\star(A) \propto \sqrt{t(A)/(\kappa(A)F(A))}$ follows. Two observations are immediate. First, holding $F$ fixed, reducing $\kappa$ one-for-one offsets the centripetal pull of $t \downarrow$ (hence diversity-enhancing policies should target $\kappa$). Second, even if $F$ is policy-invariant, $N^\star(A)$ falls with $A$ under our sign restrictions, so concentration rises monotonically with capability.

## 6. Robustness

This section tests the model's most rebuttable assumptions and shows which results survive—and under what conditions. We start by relaxing quadratic mismatch and originality costs, then allow non-uniform consumer density, move the template off the midpoint, let capability affect curvatures directly, consider partial coverage (outside option), and finally allow asymmetric marginal costs. Throughout, we preserve the basic two-stage timing (locations, then prices) and the interpretation of $A$ as an exogenous capability shifter.

*6.1 Beyond quadratic mismatch and originality costs*

**Setup R1 (general convex primitives).** Replace quadratic mismatch by a general even, twice-differentiable, strictly convex function $\varphi(\cdot)$ with $\varphi(0) = \varphi'(0) = 0$ and $\varphi''(0) > 0$. Replace the quadratic originality penalty by an even, twice-differentiable, strictly convex function $h(\cdot)$ with $h(0) = h'(0) = 0$ and $h''(0) > 0$. Preferences and costs become

$$u_i(\theta) = v - p_i - t(A)\,\varphi(\theta - x_i), K_i(x_i; A) = \kappa(A)\,h(x_i - m) + F(A).$$

Consider symmetric locations $x_1 = \frac{1}{2} - \frac{d}{2}$, $x_2 = \frac{1}{2} + \frac{d}{2}$ with $d > 0$ small enough for local approximations.

**Proposition R1.** *Under Setup R1, for $d$ in a neighborhood of 0 the unique symmetric price equilibrium has markup*

$$p^\star - c(A) = \frac{1}{2}\,t(A)\,\varphi''(0)\,d,$$

*and the location stage yields the (local) interior solution*

$$d^\star(A) = \frac{t(A)\,\varphi''(0)}{\kappa(A)\,h''(0)}.$$

*Consequently, with $\varphi''(0)$ and $h''(0)$ independent of $A$, the comparative statics of the baseline model remain unchanged: $d^\star(A)$ falls if $t'(A) < 0$ and $\kappa'(A) > 0$. $p^\star(A) = c(A) + \frac{1}{2}t(A)\varphi''(0)d^\star(A)$ falls if $c'(A) \leq 0$, $t'(A) < 0$, and $\kappa'(A) > 0$.*

**Proof.** Indifference implies, by a first-order Taylor expansion of $\varphi(\theta - x_2) - \varphi(\theta - x_1)$ around $\theta = \frac{1}{2}$, that

$$\hat{\theta} - \frac{1}{2} = \frac{p_2 - p_1}{t(A)\,\varphi''(0)\,d} + o(d).$$

Hence $\partial D_1 / \partial p_1 = -1/[t(A)\varphi''(0)d]$ and the Bertrand FOC at symmetry yields $p^\star - c(A) = (1/2)t(A)\varphi''(0)d$. Profits are

$$\Pi(d; A) = \underbrace{\left(\frac{1}{2}t(A)\varphi''(0)d\right)\frac{1}{2}}_{\text{operating profit}} - \kappa(A)\,h\left(\frac{d}{2}\right) - F(A),$$

whose derivative at $d = 0$ is $\frac{1}{4}t(A)\varphi''(0) - \frac{1}{4}\kappa(A)h''(0)d$, giving the expression for $d^\star(A)$. Monotone comparative statics follow immediately. ∎

All baseline results survive with two curvature "plugs," $\varphi''(0)$ and $h''(0)$. The sufficient statistic controlling differentiation is still a ratio. Now $t\varphi''(0)/[\kappa h''(0)]$. As long as capability scales perceived distance down and originality curvature up, homogenization and price compression go through.

*6.2 Non-uniform consumer density*

**Setup R2.** Let the consumer density be $f(\theta)$, continuous, strictly positive, and symmetric around ½ with $\int_0^1 f(\theta)\, d\theta = 1$. Keep quadratic mismatch and originality costs for clarity.

**Proposition R2.** *Under Setup R2 and symmetric locations, the indifferent boundary remains at $\hat{\theta} = \frac{1}{2} + \frac{p_2 - p_1}{2t(A)d}$. At the symmetric price,*

$$\frac{\partial D_1}{\partial p_1} = -\frac{f(1/2)}{2t(A)d},\, \frac{\partial D_1}{\partial p_2} = +\frac{f(1/2)}{2t(A)d},$$

*and the symmetric equilibrium markup becomes*

$$p^\star - c(A) = \frac{t(A)\, d}{f(1/2)}.$$

*The location FOC yields*

$$d^\star(A) = \frac{t(A)}{\kappa(A)\, f(1/2)}.$$

*Thus, as long as $f(1/2)$ does not itself vary with capability, all baseline comparative statics survive. Only the levels scale with $1/f(1/2)$.*

**Proof.** With general $f$, $D_1 = \int_0^{\hat{\theta}} f$ and $\partial D_1 / \partial p_1 = f(\hat{\theta})\, \partial \hat{\theta} / \partial p_1$. Evaluate at symmetry to obtain the slopes, then use the price FOC $D_1 + (p - c)\, \partial D_1 / \partial p_1 = 0$ with $D_1 = 1/2$. The location stage substitutes the markup and proceeds as in the baseline. ∎

A thicker density at the center (large $f(1/2)$) makes demand more sensitive and reduces markups one-for-one. It also attenuates differentiation. But the direction of all capability comparative statics remains the same.

*6.3 Moving the template and affine transformations*

**Setup R3.** Let the template be at any $m \in (0,1)$ and consider an affine change of variables $\tilde{\theta} = a\theta + b$, $\tilde{x}_i = ax_i + b$, $a > 0$.

**Proposition R3.** *The equilibrium conditions and comparative statics are invariant to shifts of the template and to positive affine re-scalings of the style space, up to the expected re-scaling of $t(A)$ and $\kappa(A)$ by $a^{-2}$ (distance-squared units).*

**Proof.** Both mismatch costs and originality costs depend on squared distances; under the affine change, squared distances scale by $a^2$. Absorb this into re-defined $\tilde{t}(A) = t(A)/a^2$ and $\tilde{\kappa}(A) = \kappa(A)/a^2$. The boundary and FOCs retain the same form, and all comparative statics in $A$ are unaffected. ∎

Where the template sits or how the design axis is scaled does not matter. Only ratios of technology parameters matter.

*6.4 Capability that changes curvatures (not only levels)*

**Setup R4.** Return to the general convex setting of Setup R1 but now allow capability to move local curvatures: $\varphi''(0; A)$ and $h''(0; A)$ may depend on $A$.

**Proposition R4.** *Locally, the differentiation choice is*

$$d^\star(A) = \frac{t(A)\,\varphi''(0; A)}{\kappa(A)\,h''(0; A)}.$$

*Hence,*

$$\frac{d}{dA} \ln d^\star = \frac{t'}{t} + \frac{\partial_A \varphi''(0; A)}{\varphi''(0; A)} - \frac{\kappa'}{\kappa} - \frac{\partial_A h''(0; A)}{h''(0; A)}.$$

*A sufficient condition for homogenization ($d^{\star\prime}(A) < 0$) is that the sum of the centripetal terms dominates:*

$$\left|\frac{t'}{t}\right| + \left|\frac{\partial_A \varphi''}{\varphi''}\right| \;>\; \frac{\kappa'}{\kappa} + \frac{\partial_A h''}{h''}.$$

**Proof.** Differentiate $\ln d^\star$. The sign condition follows immediately. ∎

If capability not only scales the distance weight but also flattens the mismatch curvature or steepens the originality curvature, those effects enter additively. The main result survives provided the *effective* centripetal pull (left side) exceeds the countervailing curvature shifts (right side).

*6.5 Partial coverage (outside option)*

**Setup R5.** Let some consumers opt out when their indirect utility is negative. Keep symmetry and assume parameter values such that the indifferent boundary between firms remains interior.

**Proposition R5.** *Fix a region of parameters in which the outside option is either slack (full coverage) or binds (partial coverage) and remains so for a neighborhood of capability levels. In either case:*

- *The price FOC can be written as $p^\star - c(A) = \Lambda(A)\,t(A)\,d$ for some $\Lambda(A) > 0$ that depends on densities at the relevant margins.*

- *The location FOC becomes $d^\star(A) = \frac{t(A)}{\kappa(A)} \cdot \Xi(A)$ with $\Xi(A) > 0$ capturing the same margin densities (equal to $1$ under uniform full coverage). If $\Lambda(A)$ and $\Xi(A)$ do not themselves trend systematically with capability (i.e., the coverage regime and margin densities remain locally stable), then all baseline comparative statics with respect to $A$ are preserved. $d^\star$ and $p^\star$ fall with capability under $t'(A) < 0$, $\kappa'(A) > 0$, $c'(A) \leq 0$. The cross-price slope strictly rises. Profits net of fixed cost strictly fall.*

**Proof (sketch).** When an outside margin binds, a unilateral price change affects demand via the interior boundary and the outside boundary. The FOC takes the generic form $D_1 + (p_1 - c)\, \partial D_1 / \partial p_1 = 0$ with $\partial D_1 / \partial p_1$ the sum of boundary-density terms. At symmetry these aggregate into a positive proportionality $\Lambda(A)/(t\,d)$, yielding the stated markup form. Substituting into the location problem produces $d^\star$ with a proportionality $\Xi(A)$. If the coverage regime does not change with $A$ locally (so the densities at the active margins are approximately constant), both $\Lambda$ and $\Xi$ can be treated as fixed multipliers, leaving signs of the $A$-comparative statics unchanged. ∎

The outside option changes the scale of markups and locations through local densities, but not the direction of the capability effects, provided capability does not push the market across the coverage boundary in the neighborhood considered.

*6.6 Asymmetric marginal costs*

**Setup R6.** Let firm $i$ have marginal cost $c_i(A) = c(A) + \delta_i$ with $\delta_1 \neq \delta_2$ but small. Originality and perceived-distance primitives remain symmetric.

**Proposition R6.** For small cost asymmetries, the location choice remains $d^\star(A) = t(A)/\kappa(A)$ to first order in $\delta_i$. The symmetric markup condition becomes

$$p_i^\star - c_i(A) = t(A)\, d^\star(A),$$

so prices differ only through costs:

$$p_1^\star - p_2^\star = c_1(A) - c_2(A) = \delta_1 - \delta_2.$$

The cross-price slope and its monotonicity in capability are unchanged.

**Proof.** Reduced-form operating profit per firm at the price equilibrium depends on the markup $t\,d$, not on the absolute level of the marginal cost. Hence, to first order in $\delta$, the location objective and FOC are unaffected, yielding the same $d^\star$. The Bertrand FOC pins individual prices one-for-one to individual costs. ∎

Small adoption or efficiency gaps shift levels of prices but leave the structure - differentiation and substitutability - untouched to first order. The centripetal comparative statics are therefore robust to moderate cost heterogeneity.

*6.7 Summary*

Across these relaxations, the core predictions are robust:

- Homogenization: equilibrium differentiation remains a decreasing function of capability whenever the perceived-distance force dominates the originality force in the generalized curvature sense.
- Price compression: equilibrium prices fall with capability under weak and standard conditions (cost pass-through plus intensified competition), including with non-uniform density and with a stable outside-option regime.
- Rising substitutability (slope): the cross-price slope rises with capability independent of costs and fixed costs. This is the most robust prediction.
- Entry tightening: net profits fall and the single-crossing entry cutoff survives as long as fixed access costs rise and gross margins shrink in the generalized ratio sense.

The only places where rates matter (rather than just signs) are when capability also shifts local curvatures of mismatch or originality, or when a change in capability flips the market between coverage regimes. Even then, simple sufficient conditions - spelled out above - preserve the main results.

## 7. Policy Implications

*7.1 A "sufficient-statistics" view for enforcers*

The model implies that a regulator can summarize static conduct and dynamic viability using two composite statistics:

Conduct statistic: $M(A) \equiv \frac{t(A)^2}{\kappa(A)}$. This is the markup component beyond cost, since $p^*(A) = c(A) + M(A)$. A fall in $M(A)$ (holding $c$ fixed) means intensified competition via homogenization. This will be observed as lower prices and higher cross-price sensitivities because $S(A) = \kappa(A)/(2t(A)^2) = 1/(2M(A))$ up to the scale factor $t(A)$.

Viability statistic (entry margin): $V(A) \equiv \frac{t(A)^2}{4\kappa(A)} - F(A) = \Pi^*(A)$. Entry (or duopoly sustainability) requires $V(A) \geq 0$. The single-crossing result implies a unique capability cutoff $A_E$ at which the market flips from viable to non-viable for a symmetric rival. If a contemplated conduct or merger shifts parameters so that $V$ turns negative, duopoly/entry is not sustainable even if prices fall.

Thus, price-only or elasticity-only screens are incomplete. Authorities should jointly track $M(A)$ for conduct and $V(A)$ for entry. The pair $(M, V)$ is directly computable from primitives that the model makes observable or calibratable.

*7.2 Market definition and elasticity screens in AI settings*

As capability increases, the model predicts rising cross-price sensitivity $S(A) = \kappa(A)/(2t(A)^2)$. SSNIP-type tools therefore tend to broaden the relevant market exactly when homogenization tightens entry. This is not a contradiction. It is a diagnostic of centripetal AI. Agencies should:

- Interpret rising elasticities cautiously. A larger cross-price elasticity is *consistent* with falling prices and rising substitutability *and yet* with lower viability $V(A)$ if $F(A)$ rises and $M(A)$ falls enough. A broadened market definition does not imply a diminished need to assess entry barriers.

- Pair SSNIP with an entry viability check. For the same products used in SSNIP, compute whether the implied change in $M$ and $F$ drives $V < 0$. The test is mechanical. Using estimated $t$ and $\kappa$ (see 7.7 below), compute $\Delta V$. If a deal makes $V$ negative, or pushes it close to zero, entry is imperiled even if the SSNIP suggests ample substitution ex post.
- Use a "Template Compression" lens. Because $d^{\backslash *}(A) = t(A)/\kappa(A)$, the observed clustering of realized outputs around a common style is itself informative about $t/\kappa$. Shrinking dispersion in outputs is a telltale for increased substitutability and for a potentially shrinking entry margin, especially if contemporaneous evidence shows rising $F$.

*7.3 Unilateral-effects analysis for mergers in creative and content ecosystems*

For a marginal change in capability or a structural change that mimics higher capability, the model implies:

$$\frac{dp^*}{dA} = c'(A) + \frac{2tt'\kappa - t^2\kappa'}{\kappa^2} < 0, \qquad \frac{d\Pi^*}{dA} = \frac{t}{2\kappa}t' - \frac{t^2}{4\kappa^2}\kappa' - F'(A) < 0.$$

Hence prices fall while profits fall, implying lower prices with higher entry barriers. A merger that reduces $t$ (stronger template pull), raises $\kappa$ (steeper originality penalty), and/or raises $F$ (stack control) will unambiguously shrink $M = \frac{t^2}{\kappa}$ and may flip $V$ negative. This is precisely the combination to which traditional price-centric screens are least sensitive.

Merger screen (model-based): require the parties to furnish a pre/post estimate of $\Delta t, \Delta \kappa, \Delta F, \Delta c$ along with evidence (contracts, technical documentation, usage data). Approve only if both:

$$\begin{cases} (i) & M_{post} \geq M_{pre} - \bar{\delta}_M \\ (ii) & V_{post} \geq \bar{\epsilon} > 0. \end{cases}$$

Here $\bar{\delta}_M$ is a small tolerance for conduct deterioration and $\bar{\epsilon}$ a positive "viability buffer." Condition (ii) ensures the deal does not move the market past the entry threshold $A_E$. Condition (i) prevents "race to the template" deals that slash variety while claiming consumer benefits through short-run price reductions.

Interpretation of cost defenses. Even if a merger yields $c'(A) \ll 0$ (large cost synergies), in this model all marginal-cost savings are competed away in the pricing stage (price-cost pass-through is one-for-one under symmetry). The net effect on profitability and entry is governed by $(t, \kappa, F)$. Thus, a valid efficiency defense must demonstrate not only reductions in $c$ but also that $\Delta \kappa \leq 0$ (not raising originality costs) and $\Delta F \leq 0$ (not raising fixed access costs). Otherwise, the deal can depress prices and depress entry, which is not pro-competitive on dynamic grounds.

*7.4 Vertical and ecosystem concerns (compute, models, templates)*

Vertical control in AI stacks maps neatly into our primitives:

- Compute/model access → $F(A)$. Exclusive access arrangements and restrictive licensing increase $F$. Remedies that lower $F$ (e.g., FRAND access to APIs/compute) relax the entry condition and push $A_E$ outward. Because $V = \frac{t^2}{4\kappa} - F$, a one-unit reduction in $F$ improves the entry margin one-for-one.

- Template governance → $\kappa(A)$. Defaults, style locks, or restrictive toolchains make deviation costlier, raising $\kappa$. Interoperability and style-transfer tools reduce $\kappa$, increasing $d^* = t/\kappa$ and improving both variety and the entry margin $V$. Because $d^*$ scales like $1/\kappa$ while $M = t^2/\kappa$, small reductions in $\kappa$ have first-order benefits on both differentiation and markup.
- User-facing standardization → $t(A)$. Bundle designs that steer creators toward universal defaults compress perceived distance ($t \downarrow$). "Plurality commitments" (multiple defaults, decentralized style packs) raise $t$ and counteract homogenization. In the model, $\partial d^*/\partial t = 1/\kappa > 0$ and $\partial p^*/\partial t = 2t/\kappa > 0$. Plural styles modestly raise prices through reduced substitutability, but they also raise variety and improve the entry margin via a larger $M$. The trade-off is explicit and quantifiable.

Vertical foreclosure test: if a platform-model owner proposes conduct that (i) increases $\kappa$ or (ii) increases $F$, require the party to show that either $M$ rises (unlikely) or that $V$ remains above a conservative buffer. Absent that showing, the conduct raises barriers while shrinking variety, even if it leaves prices unchanged or slightly lower.

*7.5 Coordinated effects and monitoring*

By construction, $S(A) = \kappa/(2t^2)$ rises with capability. Products become closer substitutes. This can affect coordinated effects in two ways. First, higher substitutability can make price responses more predictable and hence easier to monitor in a dynamic game. Second, however, margins are falling because $M = t^2/\kappa$ declines. The net implication is ambiguous. The conservative enforcement posture is to treat rising $S(A)$ as increasing the feasibility of coordination but to recognize that lower margins reduce the gain from collusion. In any case, the non-collusive channel in our model operates regardless: even with no coordination, the centripetal technology drives homogenization and entry tightening.

*7.6 Remedy design: target the right primitive*

Because outcomes are linear or rational functions of $(t, \kappa, c, F)$, remedies can be engineered to hit the bottleneck precisely:

- Lower $F$ (contestability remedy). FRAND access to compute and model APIs. Transparent, nondiscriminatory pricing. Caps on minimum-spend commitments. Each unit reduction in $F$ lifts $V$ one-for-one and does **not** distort the markup statistic $M$.
- Reduce $\kappa$ (diversity remedy). Interoperability, style-transfer toolkits, portable prompts, open stylization APIs, and unbundling of proprietary style packs. Since $d^* = t/\kappa$, this expands variety and increases $M$, counteracting entry tightening.
- Raise $t$ (plurality remedy). Mandate multiple defaults, expose style "marketplaces," or prevent exclusive template locks. This expands $d^*$ and $M$ but modestly raises prices; the model quantifies the trade-off directly via $\partial p^*/\partial t = 2t/\kappa$.
- Accept $c$ efficiencies, but do not rely on them. Lower $c$ passes through to prices one-for-one and does not improve viability $V$. A merger that only reduces $c$ is not a durable answer to variety and entry concerns created by $(t, \kappa, F)$.

Conditional approvals can be written in primitive language. Approve if $\Delta F \leq 0$ and $\Delta \kappa \leq 0$ and $\Delta t \geq 0$. Otherwise, require binding commitments that deliver equivalent parameter shifts (e.g., third-party template portability to offset an increase in $\kappa$).

*7.7 Measurement and implementation (how to estimate $t, \kappa, c, F$)*

The model is implementable with standard tools:

- Estimating $t(A)$ (perceived distance). From local demand around the symmetric equilibrium, $\partial D_1 / \partial p_2 = \kappa/(2t^2)$. If one can estimate cross-price derivatives and has a handle on $\kappa$ (see below), invert to obtain $t$. Alternatively, map observed embedding-space distances or controlled A/B style experiments into perceived differences and scale to $t$ using the pricing first-order condition $p^* - c = t\,d$.
- Estimating $\kappa(A)$ (originality curvature). Measure the incremental resource cost (compute/time/tooling) to move a design by a small $\delta$ away from the template, $\Delta K \simeq \kappa \delta^2$. With multiple $\delta$ and observed cost levels, back out $\kappa$. In practice, one can instrument for $\delta$ using exogenous "style shocks" in the toolchain or default-template rotations.
- Estimating $c(A)$. Use standard cost pass-through: $p^* - c = t\,d$ with $d^* = t/\kappa$ implies $p^* - c = t^2/\kappa$. Observing $p^*$ and estimates of $t, \kappa$ yields $c$.
- Estimating $F(A)$. Use audited fixed outlays attributable to compute/model licenses/safety/tooling and amortize over expected output. Combine with the observed $\Pi^*$ (gross margin less fixed outlays) to sanity-check $F$.

With $(t, \kappa, c, F)$ in hand, agencies can compute $(M, V)$ directly and run counterfactuals: e.g., a proposed conduct that increases $\kappa$ by $\Delta\kappa$ reduces $M$ by $t^2 \Delta\kappa/\kappa^2$ and reduces $V$ by $t^2 \Delta\kappa/(4\kappa^2)$.

*7.8 Ex ante obligations and "no-regrets" instruments*

The model points to ex ante guardrails that preserve contestability without blunting static gains:

- Homogenization stress tests. Periodically compute $M$ and $V$ for major AI ecosystems. If $V$ approaches zero, trigger remedies that reduce $\kappa$ or $F$ (e.g., interoperability mandates, FRAND compute).
- Template plurality and portability. Require a minimum number of default styles and enforce portability of user-owned style artifacts across tools. This raises $t$ and reduces effective $\kappa$.
- Access commitments. Public-option or shared-access compute reduces $F$ for non-integrated rivals, expanding the range of $A$ with viable entry.

These instruments push on $(t, \kappa, F)$ directly. These are the specific levers that matter for variety and entry in this model.

## 8. Conclusions

The single capability index $A$ moves four primitives, $t(A) \downarrow$, $\kappa(A) \uparrow$, $c(A) \downarrow$, $F(A) \uparrow$, by assumption. This delivers the core ratio results $d^* = t/\kappa$ and $p^* = c + t^2/\kappa$, and the entry margin $\pi^* = t^2/(4\kappa) - F$. But it is a reduced-form mapping. If, in a given domain, capability expands creative breadth (so $t' \geq 0$) or does not steepen originality ($\kappa' \leq 0$), homogenization and price compression may weaken or reverse. Likewise, results that compare rates rather than signs (e.g., Lerner and cross-price elasticity monotonicity) hinge on dominance conditions that are empirical. Future work should microfound $t(A)$ and $\kappa(A)$ (e.g., via attention/representation or production primitives) and test the rate restrictions directly.

The analysis is static (two-stage: location then price) with non-cooperative pricing and exogenous $A$. It abstracts from dynamic adoption, learning, and capacity or compute constraints. The adoption extension (a technology-adoption Prisoners' Dilemma) is sketched but not solved in a fully dynamic equilibrium with entry/exit. A Markov-perfect treatment that nests investment in $A$, exit, and diffusion would make the long-run structure claims tighter.

The baseline is one-dimensional Hotelling with quadratic mismatch and a quadratic originality penalty around a single "template." Robustness shows the main signs persist under general convex curvatures, non-uniform density, template shifts, partial coverage, and small asymmetries. However, the strongest welfare and elasticity claims again depend on rate conditions and on staying inside a fixed coverage regime. High-dimensional style spaces, multiple templates, and richer heterogeneity may alter both the scale and comparative statics of $d^*$.

By design, the model rules out algorithmic coordination; yet with higher substitutability, the feasibility of tacit collusion can change. The paper's mechanism is "non-collusive centripetal pressure". Coordinated-effects channels remain outside the model and could overturn price-compression predictions in some markets. A unified model that admits both channels and lets policy or monitoring constraints turn one or the other "on" is a priority.

The "template" is treated as an exogenous center. In practice, platform owners choose defaults, style packs, and access terms. Those choices are strategic and respond to regulation and rivalry. Endogenizing template governance (a leader-follower problem in $\kappa$ and $t$) would connect the theory more directly to vertical foreclosure and remedy design.

Certain extensions would therefore raise external validity, such as

- Dynamic adoption with entry/exit. Embedding the adoption sketch in a dynamic oligopoly with capacity, diffusion, and scrapping; tracking transitional concentration paths and the co-evolution of $(t, \kappa, c, F)$ with investment and policy shocks. This would turn the one-crossing entry result into a time path for viability.
- Multi-dimensional styles and plural templates. Replacing the line with $\mathbb{R}^k$ and allowing multiple locally attractive "templates." Characterizing when centripetal forces aggregate or cancel across dimensions. Deriving the mapping from portfolio dispersion to the sufficient statistic $\mu = t^2/\kappa$.
- Platform governance as choice. Let an upstream platform pick $(t, \kappa)$ through defaults, toolchains, and interoperability. Downstream firms then choose locations and prices. Solving for equilibrium plurality commitments would allow for comparison with FRAND-style remedies.
- Coverage transitions and regime switching. Formalizing the flip between full and partial coverage as a state change, endogenizing outside-option density. These would provide global conditions (beyond local multipliers) under which signs survive across regime boundaries.
- n-firm structure beyond the sketch. Closing the Salop extension with explicit location patterns, free-entry, and welfare for general $N$. Calibrating how $N^*(A)$ co-moves with $\mu$ and the entry margin $\mathcal{V} = t^2/(4\kappa) - F$.

Empirical work needs to focus on microfounding the primitives. Particular areas of priority include

- Mapping perceived distance $t$ using local demand slopes and/or controlled "style distance" experiments;

- recovering originality curvature $\kappa$ from the measured resource gradient of moving away from defaults;
- identifying pass-through to pin down $c$;
- measuring fixed access outlays $F$ from audited stacks. Then computeing $\mu = t^2/\kappa$ and $\mathcal{V} = t^2/(4\kappa) - F$ directly.

Empirical work should exploit shocks to compute/API access (FRAND openings, quota changes), template rotations, and licensing frictions to separately identify $\Delta F$, $\Delta t$, $\Delta \kappa$, and $\Delta c$. Each shock yields a sign test for the model's comparative statics without requiring full structure. Using staggered access, bans, or default changes (documented in creative tools and APIs) could trace the joint movement ($p \downarrow$, cross-price slope $\uparrow$, $d^*$, $\mathcal{V} \downarrow$) and test the rate conditions under which cross-price elasticity and the Lerner index decline. Where the model predicts a unique cutoff $\hat{A}$, searching for one-crossing patterns in establishment counts or app ecosystems as capability and fixed costs move could be useful. Relating breakpoints to observed $(t, \kappa, F)$ proxies would underpin our model. In markets with algorithmic pricing, splitting samples by enforcement or monitoring intensity (or algorithm class) could separate collusive price elevation from non-collusive centripetal compression. The model's predictions should survive in the latter.

The paper documents that the directional results survive with general convex curvatures, non-uniform density, template shifts, partial coverage (locally), and small asymmetries. This breadth is valuable. The most policy-relevant objects ($d^* \downarrow$, $p^* \downarrow$ under mild conditions, rising cross-price slopes, and a single-crossing entry margin) do not hinge on quadratic forms or uniform density. Where robustness thins is where rates matter (Lerner and elasticity monotonicity) and where regime switches occur (coverage flips). Those are the two areas where empirical work should concentrate and where theory should supply clean global conditions.

The model is falsifiable on primitives. If capability expansions are associated with $t' \geq 0$ or $\kappa' \leq 0$ in the relevant margin, the centripetal mechanism is not operative. If $\mu = t^2/\kappa$ rises while $F$ is stable, entry should not tighten even as prices fall. The practical guidance is to track $\Delta\mu$ (conduct) and $\Delta\mathcal{V}$ (viability) rather than price levels alone, and to insist on designs that isolate movements in one primitive at a time.

The central insight (that AI can drive lower prices and weaker structure simultaneously) will survive particular scrutiny if we (i) microfound and measure the primitives that generate it, (ii) follow their dynamics through investment and exit, and (iii) accommodate platform governance and coordination risks in one frame. The next iteration of the model should therefore be dynamic, multi-dimensional, and platform-aware, with an empirical program that measures $(t, \kappa, c, F)$ and stress-tests the rate conditions that make the mechanism bite.

**References.**